\documentclass{aa}

\usepackage{psfig}

\def\xuno  {LMC~X--1}

\def\xtre  {LMC~X--3}
\def\xquattro  {LMC~X--4}
\def\iras {IRAS~04575--7537}
\def\exo  {EXO~0748--676}
\def\psr {PSR~B0540--69}
\def\integral {{\it INTEGRAL}}

\begin{document}



\title{An INTEGRAL Hard X-ray Survey of the Large Magellanic Cloud
\thanks{Based on observations with INTEGRAL, an ESA project with instruments and science data centre funded by ESA member states (especially the PI countries: Denmark, France, Germany, Italy, Switzerland, Spain), Czech Republic and Poland, and with the participation of Russia and the USA.} }

   \author{D. G\"{o}tz\inst{1}, S. Mereghetti\inst{1},
    D. Merlini\inst{1,2}, L. Sidoli\inst{1}, \and T. Belloni\inst{3}
    }

   \offprints{D. G\"{o}tz, email: diego@mi.iasf.cnr.it}

   \institute{INAF -- Istituto di Astrofisica Spaziale e Fisica Cosmica,
                        Via Bassini 15, I-20133 Milano, Italy
          \and
         Diaprtimento di Fisica -- Universit\`{a} degli Studi di Milano, Via Celoria 16, I-20133 Milano, Italy
     \and
     INAF -- Osservatorio Astronomico di Brera, via E. Bianchi 46, I-23807 Merate, Italy
         }


\abstract{
Two observation campaigns performed in 2003 and 2004 with  the
\integral~ satellite have provided the first sensitive survey of the
Large Magellanic Cloud with an imaging instrument in the hard
X-ray range (15 keV - 10 MeV).
The high energy flux and long term variability of the black hole 
candidate \xuno~ was measured for the first time without the contamination 
of the nearby ($\sim$25$^{\prime}$) young pulsar \psr.
We studied the accreting pulsar \xquattro, constraining the size of the
hard X-ray emitting region ($\leq$3$\times$10$^{10}$ cm) from the analysis
of its eclipses, and measuring its spin period (13.497$\pm$0.005 s) in the 20-40 keV band.
\xtre~ was not detected, being in a soft state during the
first observation and possibly in an extremely low state in the second one.
Thanks to the large field of view of the IBIS instrument, we could study also 
other sources falling serendipitously in the
observed sky region around the LMC:  the Galactic low mass X-ray
binary \exo, the accreting pulsar SMC X-1 in the Small Magellanic
Cloud, and the Active Galactic Nucleus \iras~. In addition we discovered 
five new hard X-ray sources, two of which  most likely belong to the LMC. 
 }

\authorrunning{D. G\"{o}tz et al.}

\maketitle

\section{Introduction}

The Large Magellanic Cloud (LMC) has been extensively observed in
the soft X-ray range (E$\leq$10 keV) with imaging X--ray
satellites, which revealed a large number of X-ray sources of
different classes (see e.g. \cite{einstein} for {\it Einstein}, \cite{rosat,rosat2} for {\it ROSAT}, 
and \cite{xmm} for {\it XMM} surveys). The
brightest sources have also been observed at higher energies, but
mostly with collimated hard X-ray detectors.
Observations with non-imaging instruments include those
performed with the {\it OSO-7} satellite, which yielded only upper
limits for the LMC  point sources (\cite{oso-7}), and with the
{\it Ginga} satellite which detected \object{\xtre} (\cite{ebisawa93}),
\object{\xuno} (\cite{ebisawa}), two black hole candidates, and \object{\xquattro},
an accreting pulsar, up to
$\sim$ 30 keV (e.g. \cite{levine}). \xtre~ and \xuno~ were then studied with more
sensitive instruments like HEXTE on board {\it {\it RXTE}} (e.g.
\cite{nowak}) and the PDS on board {\it Beppo}SAX (\cite{treves}).
However, all these non-imaging instruments were affected by some
confusion problems, due to the source crowding in this region of
the sky.

A few imaging surveys of this nearby (d$\sim$ 55 kpc) galaxy in
the hard X-ray (E$\geq$10 keV) energy range were carried out, with limited
sensitivity, exploiting coded mask telescopes. Observations
with the SIGMA telescope (40-150 keV) on board the
{\it Granat} satellite failed to reveal a significant hard X-ray
flux from any of the high-energy sources of the LMC
(\cite{finoguenov}). The data taken with ART-P, on board the same
satellite, but working in a lower energy band (3-15 keV), revealed
\xuno~ and the young rotation powered pulsar \object{\psr}, without 
any significant detection above 10
keV (\cite{grebenev}). The TTM telescope (2-27 keV) 
detected only \xquattro~ and supernova SN 1987A 
above 15 keV (\cite{sunyaev,su87}).

The \integral~ observatory (\cite{integral}), thanks to the good sensitivity
and to the imaging capabilities of its instruments, 
recently performed a deep observation of the LMC in the hard-X/soft-$\gamma$-ray
domain. We will focus here on the results obtained with the ISGRI
detector layer (15 keV-1 MeV, \cite{isgri}) of the IBIS
(\cite{ibis}) coded mask telescope. Its large field of view
($\sim$29$^{\circ}\times$29$^{\circ}$) contains the entire galaxy,
and  its good angular resolution ($\sim$12$^{\prime}$) is
sufficient to resolve most of the LMC hard X-ray sources.

\section{Observations and Data Analysis}

\integral~ performed two series of observations of the LMC. The
first in January 2003, just after the end of the Performace and Verification
phase of the satellite, 
and the second in January 2004. The dates and exposure times
of the observations are reported in Table \ref{obs}.
 \begin{table}[ht!]
\caption{\integral~ LMC Observations}
\begin{center}
\begin{tabular}{ccc}
\hline
\hline
\integral       &  Orbit Start &Net exposure \\
Orbit          &   Time [UT]   & [ksec]     \\
\hline
27 & 2003-01-02 04:07:37 &  209.2 \\
28 & 2003-01-05 03:53:52 &  71.0\\
29 & 2003-01-08 03:41:54 &  212.0\\
33 & 2003-01-20 02:53:20 &  216.2\\
34 & 2003-01-23 02:38:41 &  215.1\\
35 & 2003-01-26 02:25:14 &  217.2\\
\hline
27-35 & &1140.7\\
\hline
150 & 2004-01-05 02:11:14 & 80.3\\
151 & 2004-01-08 02:00:31 & 197.3\\
152 & 2004-01-11 01:49:00 & 130.3\\
\hline
150-152 & & 407.9\\
\hline
\end{tabular}
\end{center}
\label{obs}
\end{table}
The highly inclined (52$^{\circ}$) and long ($\sim$72 hr)
orbit of \integral~ allows for long, uninterrupted observations. However, in
order to reduce the presence of the typical systematic noise of
coded mask instruments in the reconstructed sky images,
each observation  is divided  in a series of individual pointings
following a predefined pattern of sky directions around the
target. The 2003 data were obtained  in ``hexagonal dithering"
mode, which consists of the cyclic repetition of a pattern
composed by 7 pointings: the first one aimed at SN 1987A and the
following six spaced by two degrees at the vertices of a hexagon
centred on the first pointing. The 2004 observation was instead
performed using a dithering mode consisting of 25 pointings,
spaced again by 2 degrees, arranged over a 5$\times$5 square grid.

We  analysed the IBIS data using version 4.2 of the Offline
Scientific Software (OSA) provided by the \integral~ Science Data
Centre (\cite{isdc}).

\subsection{Imaging}

After standard data processing (dead time correction, good time
intervals selection, gain correction, energy reconstruction), we
produced the sky images of each individual pointing in four
energy bands: 20-40, 40-60, 60-100 and 100-150 keV. No sources are
visible in the single pointings, owing to their short duration,
with the exception of \xquattro~ (see below). To look for fainter
sources, we produced the mosaics of the single orbits ($\sim$ 200 ks) and of the two entire
observations, after flat-fielding each individual sky image using the median of all
the images of the same orbit. This procedure reduces the
systematic noise induced by the non-uniformity of the detector
shadowgrams, and thus improves the quality of the mosaic images.

\begin{figure}
\centerline{\psfig{figure=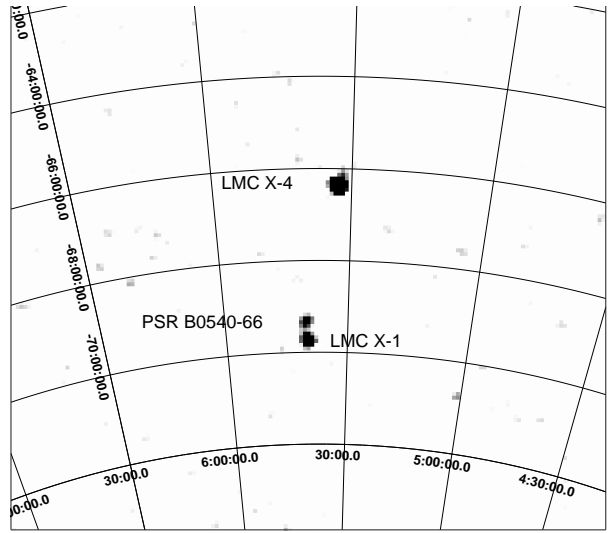,width=8cm}}
\caption{IBIS/ISGRI image (20-40 keV) of the LMC region obtained in 2003. For the first
time the flux of \xuno~ can be clearly disentangled from the one of \psr~ at energies above 20 keV.}
\label{2003img}
\end{figure}

In the 20-40 keV mosaic of the 2003 observation we detected three
LMC sources,  \xuno, \xquattro~ and \psr~ (see Fig. \ref{2003img}),
the Galactic low mass X-ray binary \object{\exo} and the Seyfert 2 galaxy
\object{\iras}.
In the 2004 mosaic in the same energy band  we detected \xquattro, 
\psr , \exo , and, at a large off-axis angle also a source in
the Small Magellanic Cloud, \object{SMC X--1}. Due to source variability
\xuno~ was below the sensitivity threshold for the 2004
observation, which was too short to detect \iras. 

In addition to these known objects, we discovered  a few new hard
X-ray  sources in the 20-40 keV band. They were detected  in the partial and total
mosaics as indicated in Table \ref{newsrc}. To evaluate the
significance of their detection, we scaled the significance
computed by the OSA software, to the $\sigma$ values derived by
fitting a Gaussian to the pixel distribution of the final
significance images. The coordinates, detection significance and
fluxes of the new sources are reported in Table \ref{newsrc}. The   uncertainty on
their coordinates is about 3.5$^{\prime}$.
\begin{table*}[ht!]
\caption{New sources detected during the LMC observation campaign}
\begin{center}
\begin{tabular}{ccccccc}
\hline
\hline
Name (IGR) & R.A. (J2000) & Dec. (J2000) & 2003 $\sigma$ & 2003 Flux [mCrab]  & 2004 $\sigma$  & 2004 Flux [mCrab]\\
& $hh:mm:ss$ &$^{\circ}~:~^{\prime}~:~^{\prime\prime}$ & & 20-40 keV & & 20-40 keV\\
\hline
J03532--6829 & 03:53:14 & --68:29:00 & 3.9 & 0.6 & - &- \\
J05009--7044 & 05:00:59 & --70:44:45 & 4.1$^{1}$ & 0.7 & - & -\\
J05319--6601 & 05:31:57 & --66:01:35 & 6.2 & 0.9 &- &-\\
J05346--5759 & 05:34:35 & --57:59:35 & 4.4 & 1.2 & - &-  \\
J06239--6052 & 06:23:58 & --60:52:15 & - & - & 4.6 & 1.8 \\
\hline
\end{tabular}
\end{center}
$^{1}$ reaches 5.5 $\sigma$ (2.5 mCrab) during orbit 27.
\label{newsrc}
\end{table*}

\subsection{Spectral Extraction}
\label{spectra}

To derive the average fluxes and spectra of each source
we used the count rate values obtained from the mosaic sky
images and rebinned the IBIS/ISGRI response matrix in order to match
the four energy channels we used. 
This spectral extraction method is indicated for weak
sources and has already been used successfully (see e.g. \cite{sgr}).
In addition, for the brightest sources (\xquattro,
\xuno~(only 2003 obs.) and \exo), we performed the imaging analysis
and derived the spectra also in nine energy bands between 20 and
200 keV.  The spectra were fitted with XSPEC v 11.2
(\cite{xspec}). A systematic error of 10\% has been added to the
data, to account for the uncertainties of our spectral
extraction method and of the response matrix.  All the
spectra were fitted with a power law model, with the
exception of \xquattro~ and \exo, for which a cutoff power law
($E^{-\Gamma}\times e^{(-E/E_{c})}$, where $E_{c}$ is the cutoff energy 
and $\Gamma$ the photon index)
gives  better results. The photon spectra have then been
multiplied by a constant factor ($\sim$1.7) derived from the
comparison with observations of the Crab Nebula.

In order to look for spectral variability, we also considered
the mosaics of the single orbits. With
the exception of  \xquattro~ (see below), all sources were
consistent with a constant flux within the same observation,
while, as mentioned above, some long term variations appeared by
comparing the fluxes in 2003 and 2004 (\xuno).
All the fluxes and spectral parameters, referring to the 20-100
keV energy range, are reported in Table \ref{fluxtab}.
\begin{table*}[ht!]
\caption{Fluxes (20-100 keV) and spectral parameters derived for the sources detected by IBIS/ISGRI during
the 2003 (upper panel) and 2004 (lower panel) observations. The errors are all at 1 $\sigma$ level for
one parameter of interest.}
\begin{center}
\begin{tabular}{c|c|c|c|c|c}
 \hline\hline
 & \xuno & \xquattro & \psr & \exo & \iras \\
 Flux [10$^{-12}$ erg cm$^{-2}$ s$^{-1}$]& 58$^{+8}_{-6}$ & 124$\pm$60 & 31$^{+14}_{-10}$ & 320$\pm$150 & 15.3$\pm$5.1 \\
 Photon Index $\Gamma$& 2.9$\pm$0.3 & 1.0$\pm$0.2 & 2.2$\pm$0.4 & 1.3$\pm$0.4 & 3.4$\pm$1.0 \\
 Cutoff Energy $E_{c}$ [keV] & - & 13$\pm$5 & - & 44$\pm$15 & - \\
 \hline
 Flux [10$^{-12}$ erg cm$^{-2}$ s$^{-1}$]& - & 590$\pm$100 & 24$\pm$7 & 421$\pm$150 & - \\
 Photon Index $\Gamma$ & - & 1.35 $\pm$ 0.2 & 1.9$\pm$0.5 & 1.6$\pm$0.4 & - \\
 Cutoff Energy $E_{c}$ [keV] & - & 17 $\pm$ 7& - & 50$\pm$15 & - \\
 \hline
\end{tabular}

\label{fluxtab}
\end{center}
\end{table*}

\subsection{Timing analysis}

\subsubsection{\xquattro}

\begin{figure}
\centerline{\psfig{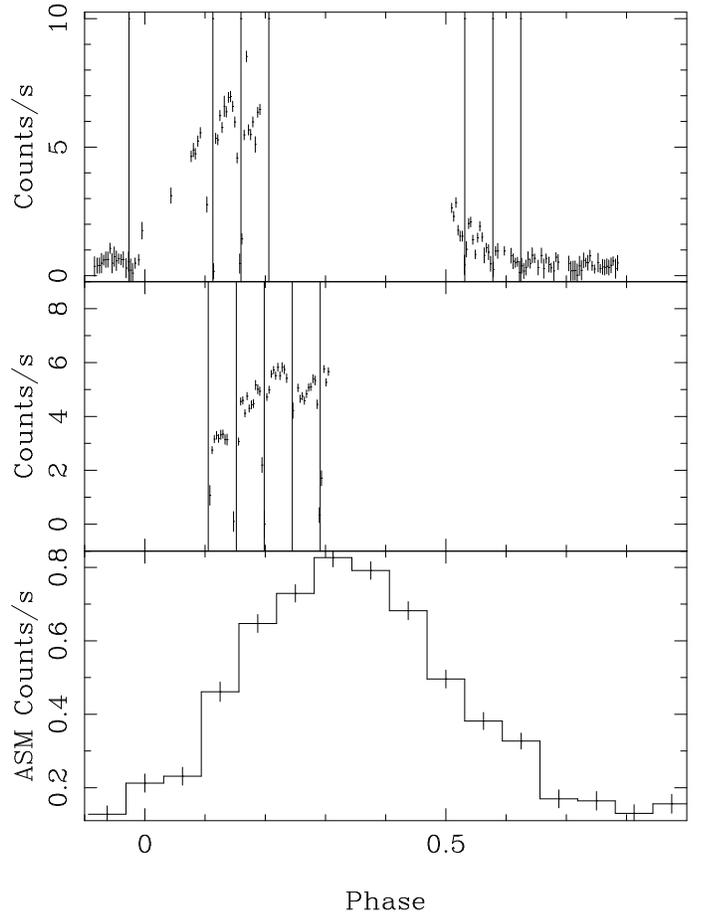}}
\caption{Upper panel: IBIS/ISGRI light curve of LMC X-4 during the
2003 observation (20-40 keV). The time has been converted to super-orbital phase
(period=30.3 days). 
The original data have been rebinned by a factor 4 ($\sim$8800 s time bins).
Middle panel: IBIS/ISGRI light
curve of LMC X-4 during the 2004 observation (20-40 keV).  The vertical lines
represent the times of mid eclipse as derived by the ephemeris of
\cite{naik}. Lower Panel: RXTE/ASM light curve of LMC X-4 folded
at the super-orbital period. For all plots phase=0 corresponds to 52644 MJD.} \label{x4lc}
\end{figure}

From the mosaics of the individual orbits we noticed that the flux
of \xquattro~ increased by a factor $\sim$7 in 6 days during the
first part of the January 2003 observation, and returned to the 
initial level towards the end
of the observation. This
can be clearly seen from the light curve obtained
from the fluxes measured in the individual pointings ($\sim$ 2200 s each, Fig. \ref{x4lc}, upper panel).
The source variability  can be attributed to  the
30.3-day super-orbital period observed at lower energies (e.g.
\cite{paul}) and generally interpreted as the effect of precession
of a tilted accretion disk. The bottom panel of Fig. \ref{x4lc}
shows for comparison the 2-10 keV light curve obtained by
folding at $P_{S}$=30.296 days the data of the RXTE ASM
instrument.
Our 2003 observation, extending over 26 days, covered
most of the super-orbital period but missed the peak of the source
intensity (top panel of Fig. \ref{x4lc}), while a much shorter phase
coverage was obtained in 2004 (Fig. \ref{x4lc} middle panel).

Referring to the upper panel of Fig. \ref{x4lc}, we define high and low state
of the source when the 20-40 keV flux is above/below 4 counts/s.
The spectra extracted over the two states differ
only in flux (from 7$\times$10$^{-11}$ to 4.3$\times$10$^{-10}$
erg cm$^{-2}$ s$^{-1}$) but show a similar shape. For comparison we have extracted
the high state spectra of \xquattro~ also using the standard OSA
4.2 spectral extraction, which is suitable for bright sources that
are clearly detected during single pointings. In fact \xquattro~
was detected also in the single pointings in this case. The two
methods gave similar results.

When the source was in the high state, the  eclipses due to its
occultation  by the giant companion star ($P_{orb}$= 1.4084 days,
\cite{white}) are visible in the \integral~ light curves.
To study the eclipses
in more detail we extracted the light curves (20-40 keV)  with 100
s resolution around the time of the eclipses, both for the 2003
and 2004 observations.
\begin{figure}
\centerline{\psfig{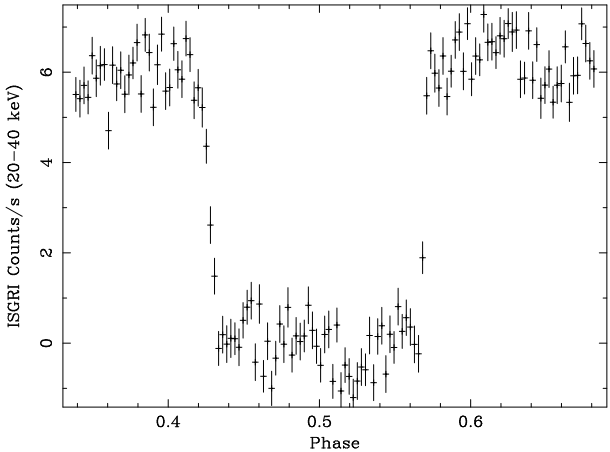}}
\caption{IBIS/ISGRI (20-40 keV) average light curve of the 
two eclipses of \xquattro~ used in our analysis (see text). Time bin = 300 s.}
\label{x4ecl}
\end{figure}
We added in phase the data of the two complete eclipses of the
2004 observation, see Fig. \ref{x4ecl}. 
The eclipse ingress seems to be resolved and  
in fact it can not be fitted with a step function. The same is not true for the egress.
Assuming that the ingress is truly resolved, we can derive some information on the size 
of the largest hard X-ray emitting region.
If the geometry of the binary system and its projection on the sky 
are known, we can assume that at the beginning of the eclipse
the projection of the distance between the
centres of the two binary components coincides with the sum of their
radii, while at the beginning of the total occultation it coincides
with their difference. We indicate with $R_{S}$ and $R_{X}$ the radii
of the companion star and the hard X-ray emission region respectively, with
$R$ the binary separation (8.5$\times$10$^{11}$ cm, \cite{levine00}) and with $\theta = 2\pi\phi$ the phase ($\theta_{1}$ for the
early beginning and $\theta_{2}$ for the start of the total occultation and obtain the
following relations:
\begin{equation}
\label{eq1}
R_{X}+R_{S}=R(\sin^{2}\theta_{1}+\cos^{2}\theta_{1}\cos^{2}i)^{1/2},
\end{equation}
\begin{equation}
\label{eq2}
R_{X}=R_{S}-R(\sin^{2}\theta_{2}+\cos^{2}\theta_{2}\cos^{2}i)^{1/2}
\end{equation}

$\theta=0$ corresponds to the conjunction of the two stars with the compact object
closer to the observer, $\theta=\pi$ corresponds to the mideclipse time.
By fitting the light curve with a constant plus an exponential cutoff, 
one can derive a precise value of $\theta_{1}=2.614$ and $\theta_{2}=2.714$.
Substituting Eq. \ref{eq1} in Eq. \ref{eq2} and knowing the geometry
of the system (inclination angle $i$=68$^{\circ}$, \cite{levine})
we derive a size for the high energy emitting region, $R_{X}$, 2.63$\times$10$^{10}$ cm.

\begin{figure}
\centerline{\psfig{figure=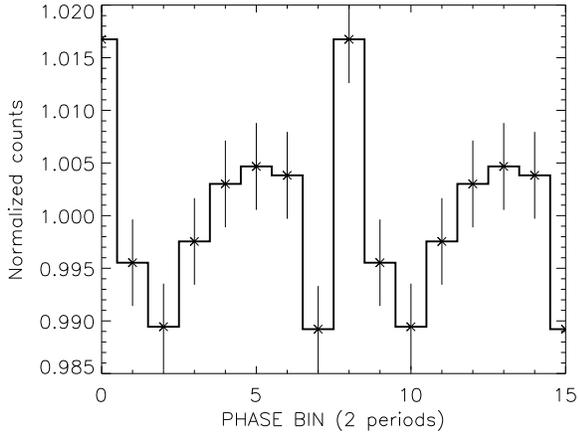,width=9cm,angle=0}}
\caption{\xquattro~ 20-40 keV light curve folded at the best fit period of 13.497$\pm$0.005 s.}
\label{x4fold}
\end{figure}

We  looked for the presence of the $\sim$13.5 s neutron star spin
period
by analysing the pointings during which the source was brighter.
The photon arrival times were converted to the solar system
barycentre and corrected for the orbital motion of the system. We restricted
our search to the 15-40 keV energy band, the one with the highest
signal-to-noise ratio. An additional selection criterion to
improve our statistics was to chose only the pixels illuminated by
the source for at least 75\% of their surface. We applied an epoch
folding analysis
to search in the range from 13.4 s to 13.6 seconds, with 8 phase
bins and 150 trial periods. This period range is much larger than
the range of periods ever observed in this source (see Fig. 2 in
\cite{woo}).
We could not detect the pulsations in the 2003 data, which, having
been taken at the beginning of the mission, when the on board
parameters of the instruments were not yet completely  optimised,
are affected by a higher background noise. The search was more
fruitful for the 2004 observation, for which pulsations were
detected considering together five time series taken from orbit 152, starting at
53016.54105 MJD and ending at 53016.71887 MJD (total exposure 11.5
ks). Fitting the distribution of  $\chi^{2}$ values versus trial
period as described in \cite{leahy}, we
obtained a best-fit period of 13.497$\pm$0.005 s. The folded light curve 
(Fig. \ref{x4fold}) shows two peaks per phase, in contrast
to the single broad peak measured between 15 and 37 keV by {\it Ginga}
(\cite{woo}), and with SAX between 23 and 30 keV (\cite{naik}).

\section{Discussion}

\subsection{\xuno}

\xuno~ is a well known black hole candidate. Together with Cyg X-1
and \xtre, it is the only persistent X-ray binary with an
accurately determined mass function strongly suggesting the
presence of an accreting black hole (\cite{lmcx1mass}). \xuno~ spends most
of the time in the so called  high/soft state, characterised by a
spectrum modeled with a multi-temperature disk black body plus
a power law component dominating above 
10 keV and a narrow emission line at 6.5 keV. 
The photon index of the power law component is usually around 3 (\cite{nowak}).

\begin{figure}
\centerline{\psfig{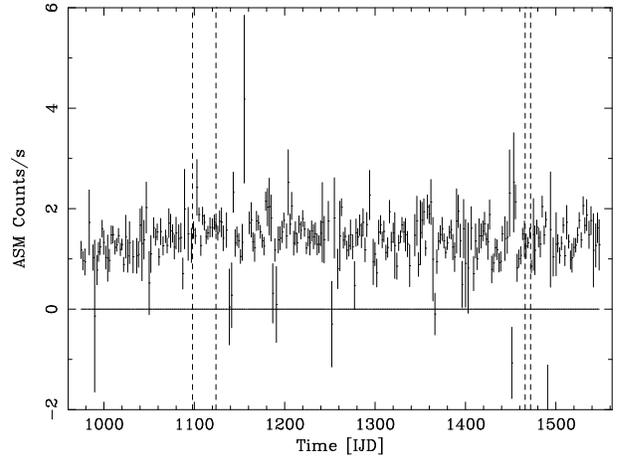}}
\caption{ASM light curve (2 days-averages) of LMC X-1. The two \integral~ observations
occurred in the periods delimited by vertical dashed lines.}
\label{asmx1}
\end{figure}

This is the first time that the high-energy
flux of this source can be fully disentangled (see Fig. \ref{2003img}) from that of \psr,
which lies just 24.6$^{\prime}$ from it (see e.g. \cite{haardt}).
We detected \xuno~  up to $\sim$80 keV only during the 2003
bservation.  
The flux measured in 2003 
(see Table \ref{fluxtab}) was one order of
magnitude higher than that measured by {\it SAX} in the same
energy range (\cite{haardt}) and a factor 5 larger than that reported with {\it RXTE}
(\cite{nowak}). This could be interpreted as a transition
of the source to its low/hard state. However the photon index ($\sim$3) 
and the fact that the 2-10 keV flux, measured
with the All Sky Monitor (ASM) on board {\it RXTE} (see Fig. \ref{asmx1}),
remained constant,
indicate that the source was in high/soft state. Hence the 2003 observation
represents the highest hard X-ray flux reported so far for the soft state of \xuno.

The non-detection of \xuno~ in 2004 indicates that
it became fainter by a factor $\sim$10.
The 3$\sigma$ upper limit on its flux in
January 2004 is 2.6$\times$10$^{-4}$ ph cm$^{-2}$ s$^{-1}$ in the
20-100 keV band (assuming the same spectral shape
of the 2003 observation).  
This value is still compatible with the {\it SAX}/PDS measurement. 
The 2-10 keV ASM
flux remained almost constant during both observations (see Fig.
\ref{asmx1}), indicating we are again measuring a soft state of the source. 
A similar behaviour, with large variations of the hard-X
ray component independent of the soft component, has been observed 
in this source and in \xtre~ using {\it Ginga} data (\cite{ebisawa,ebisawa93}). 
This seems to indicate that there is a mechanism responsible for the variation of
high-energy emission, which does not involve variations of the low-energy
flux.

\subsection{\xtre}

\xtre~ is another black hole candidate, similar in many respects
to \xuno~ (\cite{nowak}), with the main difference that it shows
a larger degree of variability at soft X-rays. In fact it undergoes
strong variations in luminosity associated with the cyclic transitions
between its low/hard and its high/soft state. During
the low/hard state the flux dims at low energies, but grows
at high energies, pivoting at $\sim$10 keV. The spectrum can then be represented
by a single power law with photon index $\sim$1.7 (\cite{boyd,wilms}). 

\begin{figure}
\centerline{\psfig{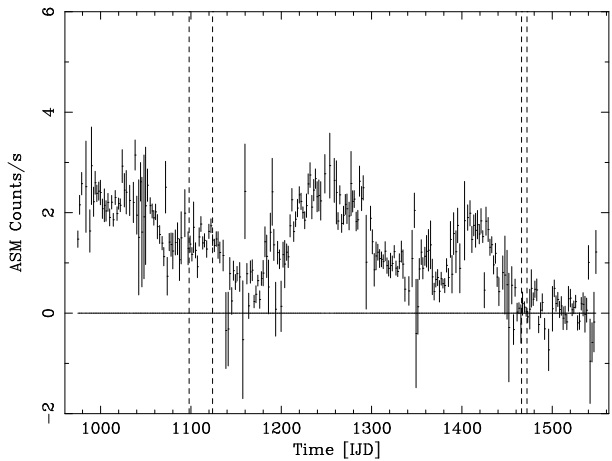}}
\caption{ASM light curve (2 days-averages) of LMC X-3. The two \integral~ observations
occurred in the periods delimited by vertical dashed lines.}
\label{asmx3}
\end{figure}

\xtre~ was  not detected in our \integral~ observations. We can put a 3$\sigma$ upper limit on
its flux (20-100 keV) of 1.8$\times$10$^{-4}$ ph cm$^{-2}$
s$^{-1}$ deduced from the first observation. Following Wilms et al. (2001) one
can associate the minima of the ASM light curve with the hard states
of the source. By comparing our data with the ASM (see Fig. \ref{asmx3}), we can conclude
that we did not observe in 2003 the hard state of the source.
In fact our first observation was close but not exactly corresponding
to a minimum. Our non-detection can be explained with the source
being in a soft state without a strong high energy component similarly
to what we observe during the second \xuno~ observation and to what reported
for the March 1988 {\it Ginga} observation (\cite{ebisawa93}). 

The second \integral~ observation, on the other hand, corresponds
to an extremely low X-ray state (no detection in the ASM light curve), 
close to quiescence, similar to the ``off" states of transient black hole binaries.
This picture is consistent with our upper limit ($\sim$3$\times$10$^{-4}$ ph cm$^{-2}$)
derived for the 2004 observation, which is lower than the extrapolation of the flux measured
at lower energies with the PCA during low/hard states (\cite{boyd}).

\subsection{\xquattro}
LMC X-4 is a high-mass X-ray binary system 
composed by an O/B type star (\cite{sandu}), probably a main-sequence star (\cite{woo}), and a pulsating neutron star, with an intense magnetic field (\cite{labarbera}), accreting material from the companion by stellar wind. Like for Her X--1, an accretion disk 
thicker in the outer regions, truncated at 
the Alf\'en radius by the magnetic pressure, is also present (\cite{lang}).

The disk emits thermal photons which contribute to the main soft (0.1-2 keV) component of the spectrum. 
Hard X-ray photons are instead produced by reprocessing in an external hot corona, that surrounds the inner part of the disk: soft photons are, in general, Compton-inverse upscattered to higher energy. When the disk is edge-on,  only the contribution due to the corona (\cite{woo}) is visible. This behaviour is confirmed by the absence of spectral variations between high and low state (see also \cite{naik03,naik}). 

If the indication of the resolved ingress duration derived from our data is confirmed, the corresponding size for the region emitting above 20 keV is
clearly not compatible with the size of the neutron star,
confirming that the hard X-rays are produced in a hot
scattering corona surrounding the accretion disk. 

\xquattro~ is thought to be near the condition of spin equilibrium, which occurs when 
the spin period of the neutron star is close to the 
Keplerian period of the inner boundary of the accretion disk.
This condition is reflected in a rather constant spin period evolution.
In fact the latter has been studied for more 
than 20 years (see e.g. \cite{woo,levine00,naik}) and has shown some variations 
but no particular spin-up or spin-down trend. 
Our measurement is still in the range of the reported periods for this source.

\subsection{\psr}
\label{psr}

\psr~ is one of the youngest rotation-powered pulsars. It was
discovered in the X-rays by the {\it Einstein} Observatory
(\cite{seward}). Its rotational characteristics (P=50 ms and
$\dot{P}$ = 4.79$\times$10$^{-13}$ s s$^{-1}$) resemble those of
the Crab pulsar and give a characteristic age of
$\sim$1.6$\times10^{3}$ yr. Like the Crab, it is surrounded by a
bright radio, optical and X-ray synchrotron  nebula.

The source was detected in our data up to 60 keV with a
marginal ($\sim$ 3$\sigma$) detection in the 60-100 keV band. 
This is the first time that a measure of the
total flux from this pulsar and its nebula can be done at these
energies. Up to now only the pulsed component could be measured
with {\it RXTE} (\cite{deplaa}) up to 48 keV.

\begin{figure}
\centerline{\psfig{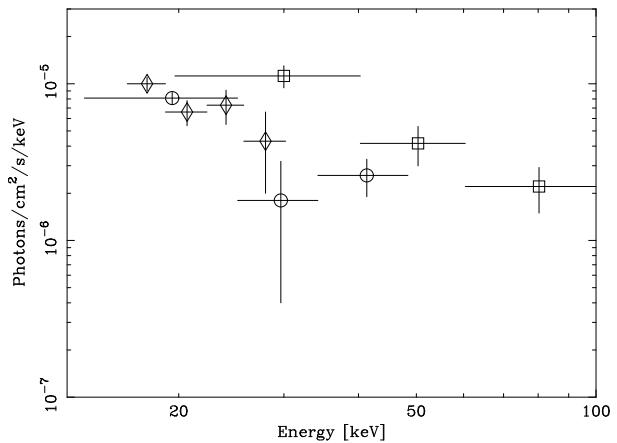}}
\caption{\psr~ spectrum above 15 keV. The squares represent the total
flux derived here with IBIS/ISGRI. The circles and the diamonds represent
the pulsed emission measured with {\it RXTE}/HEXTE and {\it RXTE}/PCA 
respectively (\cite{deplaa}).}
\label{0540}
\end{figure}

The flux we measure is 
about three times higher than the pulsed one measured by {\it RXTE}, see Fig. \ref{0540}.
This means that the nebular and the unpulsed contributions at these energies would
then be of the order of 50-75\% of the total flux.  
This can be compared
to what is reported in the same energy range for the Crab nebula, for which the
nebula and unpulsed flux represent the 85\% of the total flux (\cite{kuiper}). 

The spectral index we derive is compatible, within the errors, with
the one measured with {\it RXTE}.
A softening at higher energies has been reported for the HEXTE data on
the pulsed emission: a spectral break in our data cannot be
ruled out, due to the low statistics. 

\subsection{\exo}

The low-mass X-ray binary (LMXB) \exo, discovered with {\it
EXOSAT} (\cite{parmar}), has been extensively studied in the
soft X-ray band (0.1-10 keV). It is a peculiar source, showing all
types of temporal variability typical of the different kinds of LMXBs: 
it shows eclipses ($P_{orb}$=3.82 hr), dips, flares, and type
I X-ray bursts indicating the presence of a neutron star in the
system. 
The estimated distance of this source, lying outside the Galactic
Plane at $b\simeq -$20$^{\circ}$, is $\sim$10 kpc, and the mass of
the companion is estimated to be $\sim$ 0.5 M$_{\odot}$
(\cite{gottwald,parmar2}). 
A physical model for \exo~ was proposed by \cite{cb}: the blackbody emission
seen at low energies originates from the neutron star and an
extended Comptonized emission from the Accretion Disc Corona (ADC)
causes the cutoff power-law component measured in the hard X-ray
band. Recent XMM-Newton results based on the absence (\cite{bb})
or presence (\cite{homan03}) of eclipses in the soft energy band
(E$<$ 2 keV), seem to indicate that the dimensions of the
Comptonized region vary and, in general, that the source
has different states reflecting different geometries of
the system.

The only hard X-ray observation of this source was performed by
{\it Beppo}SAX in October 1998. The source was detected in
the PDS up to $\sim$100 keV (\cite{sidoli}). A broad-band (0.1
keV-100 keV) persistent (non-dip, non-burst) spectrum was
obtained. They derived a 20-100 keV flux of 3.68$\times$10$^{-10}$
erg cm$^{-2}$ s$^{-1}$ which is fully compatible with the flux
derived here. Also the cutoff energy of 48 keV found in the SAX
observation agrees with our data. This indicates, that despite the
large flux variations found in the soft X-rays, the hard component
seems to be more stable.

\begin{figure}
\centerline{\psfig{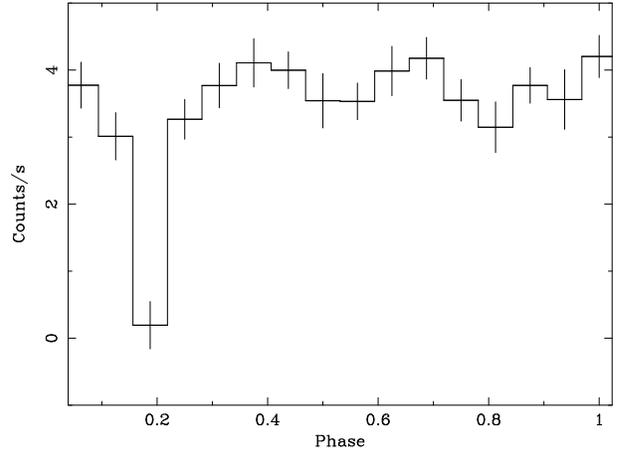}}
\caption{IBIS/ISGRI light curve of \exo~ (20-100 keV) folded at
the orbital period. The count rate is corrected for vignetting.}
\label{0748ecl}
\end{figure}

We extracted the light curve (20-100 keV) of \exo~ with 600 s time bins
(corresponding roughly to the eclipse duration at lower energies) 
and folded it at the orbital period value. The
resulting light curve, Fig. \ref{0748ecl}, shows a clear
indication of the eclipse. However, the low statistics due to the
fact that \exo~ was detected at a very large off-axis angle, does
not allow us to perform more detailed analysis.
The eclipse above 20 keV is consistent with being total, indicating that the
angular size of the companion star is greater than that of the hard X-ray emission
region.

\subsection{\iras}

The AGN \iras~ is associated with a rather bright barred spiral
Galaxy at redshift z=0.0184 and showing strong infrared
emission (\cite{hew}). \iras~ was first detected in the X-ray
range with HEAO-1, and later observed with {\it ROSAT}, {\it
Ginga} and {\it ASCA} (\cite{vignali}). It is classified as a
Seyfert 2 galaxy.

In our observations it was quite faint ($\sim$3 mCrab) and
detected it up to $\sim$80 keV
in the IBIS/ISGRI mosaic.  We derived two spectral points and an
upper limit, which represent the first measurement of this AGN at
these energies. The flux is about 1.5$\times10^{-11}$ erg
cm$^{-2}$ s$^{-1}$ (20-100 keV). This implies a 20-100 keV luminosity of
1.2$\times10^{43}$ erg s$^{-1}$ ($H_{0}$=70, $\lambda_{0}$=0.7).

\subsection{New sources}
\object{IGR J05319--6601} is located at a
position consistent with the LMC. It is at an angular distance of
21$^{\prime}$ from \xquattro,  and their point spread functions
partially overlap, therefore its significance and flux are likely to be
influenced by the brighter source. Within its error box ($\sim$ 3.5$^{\prime}$) there are
at least 7 faint unidentified soft X-ray sources, discovered during different
X-ray surveys of the LMC. Hence we cannot exclude
that we are measuring the blended emission from different unresolved
sources. The 20-40 keV luminosity we derive for this source,
supposing it is located in the LMC, is of the order of 10$^{36}$ erg s$^{-1}$.
This value and the fact that IGR J05319--6601 is close to the centre of the galaxy suggest
that we are dealing with a binary system.

\object{IGR J05009--7044}  is possibly the same source recently
reported as \object{IGR J05007--7047} and for which a {\it Chandra} X-ray
counterpart has been found (\cite{halpern}). The {\it Chandra}
source is optically identified with the bright blue star 
\object{USNO-B1.00192-00057570} (\cite{monet}). This suggests that IGR J05009--7044
could be a High Mass X-ray Binary (HMXB) in the LMC. In this case it
would have a 20-40 keV luminosity of $\sim$7$\times$10$^{35}$ erg
s$^{-1}$. However, inside the  error box of IGR J05009--7044 there
is also a radio emitting galaxy (\object{SUMSS J050059--704225}, $\sim$10
mJy at 843 Mhz, \cite{sumss}), so we cannot exclude that the
INTEGRAL source is instead an AGN. 

\begin{figure}[ht!]
\centerline{\psfig{figure=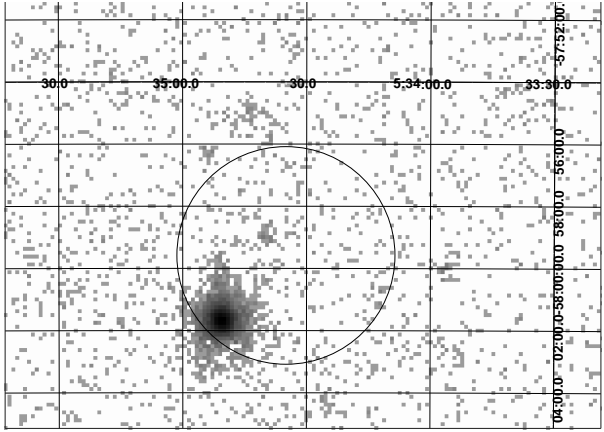,width=8cm}}
\caption{{\it ROSAT}/PSPC 0.2-2 keV image of TW Pic. The circle represents the error
box associated with IGR J05346--5759.}
\label{errbox}
\end{figure}

All the other new sources we discovered are at positions not consistent with the LMC. 
\object{IGR J05346--5759} is spatially coincident with \object{TW Pic},
see Fig. \ref{errbox}, a cataclysmic 
variable of the DQ Her type (\cite{twpic}). 
We analysed an archival {\it ROSAT}/PSPC observation of this source. 
The observation is $\sim$ 10 ks long and was performed on April 10 1992. The source
spectrum is well fitted with an absorbed power law model. We derive a photon index
$\Gamma$= 1.3$\pm$0.2 and a hydrogen column density $N_{H}$=(1.2$\pm$0.5)$\times$10$^{20}$ cm$^{-2}$,
with an unabsorbed 0.5-2 keV flux of 3$\times$10$^{-12}$ erg cm$^{-2}$ s$^{-1}$. The extrapolation
of this spectrum to the 20-40 keV energy band yields a flux of 1.3$\times$10$^{-11}$ erg cm$^{-2}$ s$^{-1}$,
which is fully compatible with the $\sim$1$\times$10$^{-11}$ erg cm$^{-2}$ s$^{-1}$ value derived from the
IBIS data. 

\object{IGR J03532--6829} can be associated with
\object{1RXS J035257.7--683120}, which has been discovered in the X-rays by the
{\it Einstein} satellite (\cite{cluster}) and detected also with {\it RXTE} in the
8-20 keV band (\cite{cluster2}) and is likely associated with the AGN
\object{PKS 0352--686}. 
\object{IGR J06239--6052} is coincident with a radio galaxy
(\object{SUMSS J062346--605222}, \cite{sumss}), making it an AGN candidate.

\section{Conclusions}
The \integral~ survey of the LMC allowed us for the first
time to study  the hard X-ray emission of some of the sources of this galaxy. 
Among the sources of the LMC we detected and studied \xuno, \xquattro~ and \psr.
\xtre~ was not detected during our observations, being in its soft state
during the first one and close to quiescence in the second one.

For the first time we were able to disentangle the high-energy emission
of \psr, for which we measured the total (pulsar+nebula) spectrum,
from the one of \xuno, for which we measured the flux and long
term variability.

Our data on the eclipses of the accreting X-ray pulsar \xquattro~ suggest that the size of the high-energy emission region is $\sim$3$\times$10$^{10}$cm.

Several serendipitous sources were detected during our survey: the brightest ones
are \exo, \iras~ and SMC X--1. We derived their fluxes and spectral
parameters, where possible. The eclipses of \exo~ were reported 
for the first time at energies above $\sim$10 keV. 

In addition we reported the discovery of 5 new faint high-energy sources. Two of them
are most likely associated with AGNs. One represents the first high-energy detection of
the cataclysmic variable TW Pic and the last two are potentially located in the LMC:
One (IGR J05319--6601) has no clear association and could be due to the blended emission of several sources and the other (IGR J05009--7044) may be an HMXB.

\begin{acknowledgements}
We acknowledge the Italian Space Agency financial and programmatic support
via contract I/R/046/04.

\end{acknowledgements}

\end{document}